\documentclass[prd,aps, reprint, nolongbibliography,twocolumn,amsfonts,amssymb,amsmath,showpacs,letterpaper,superscriptaddress,nofootinbib,altaffilletter]{revtex4-2}
\usepackage[citecolor=blue,linkcolor=blue,colorlinks=true,urlcolor=blue]{hyperref}

\usepackage{multirow}

\usepackage{graphicx}
\usepackage{amssymb}
\usepackage{amsmath}
\usepackage{longtable}
\usepackage{rotating}
\usepackage{color}
\usepackage{epsf}
\usepackage{fancyhdr}
\usepackage{ifthen}
\usepackage{slashed}
\usepackage{xspace}
\usepackage{tabularx}

\newcommand{\pastro}{$P_{\mathrm{astro}}$\xspace}

\begin{document}

\title{Search for binary black hole mergers in the third observing run of Advanced LIGO-Virgo using coherent WaveBurst enhanced with machine learning}

\author{T.~Mishra}
\author{B.~O'Brien}
\author{M.~Szczepa\'nczyk}
\affiliation{Department of Physics, University of Florida, PO Box 118440, Gainesville, FL 32611-8440, USA}
\author{G.~Vedovato}
%\affiliation{Universit\`a di Padova, Dipartimento di Fisica e Astronomia, I-35131 Padova, Italy }
\affiliation{INFN, Sezione di Padova, I-35131 Padova, Italy }

\author{S.~Bhaumik}
\author{V.~Gayathri}
\affiliation{Department of Physics, University of Florida, PO Box 118440, Gainesville, FL 32611-8440, USA}
\author{G.~Prodi}
\affiliation{Universit\`a di Trento, Dipartimento di Matematica, I-38123 Povo, Trento, Italy}
\affiliation{INFN, Trento Institute for Fundamental Physics and Applications, I-38123 Povo, Trento, Italy}
\author{F.~Salemi}
\affiliation{Universit\`a di Trento, Dipartimento di Fisica, I-38123 Povo, Trento, Italy}
\affiliation{INFN, Trento Institute for Fundamental Physics and Applications, I-38123 Povo, Trento, Italy}
\author{E.~Milotti}
\affiliation{Universit\`a di Trieste, Dipartimento di Fisica, I-34127 Trieste, Italy}
\affiliation{INFN, Sezione di Trieste, I-34127 Trieste, Italy}
\author{I.~Bartos}
\author{S.~Klimenko}
\affiliation{Department of Physics, University of Florida, PO Box 118440, Gainesville, FL 32611-8440, USA}

\begin{abstract}
%The Coherent WaveBurst (cWB) search algorithm identifies generic gravitational wave (GW) signals in the LIGO-Virgo data by looking for excess power events in the time-frequency domain, with minimal assumptions on the signal model. Since the publication of GWTC-2~\cite{GWTC2}, we have improved the cWB search sensitivity to the particular class of GW signals known as binary black hole (BBH) mergers by using a Machine Learning (ML) method~\cite{MLPaper}. 
In this work, we use the coherent WaveBurst (cWB) pipeline enhanced with machine learning (ML) to search for binary black hole (BBH) mergers in the Advanced LIGO-Virgo strain data from the third observing run (O3).
We detect, with equivalent or higher significance, all gravitational-wave (GW) events previously reported by the standard cWB search for BBH mergers in the third GW Transient Catalog (GWTC-3). The ML-enhanced cWB search identifies five additional GW candidate events from the catalog that were previously missed by the standard cWB search. Moreover, we identify three marginal candidate events not listed in GWTC-3. %, and estimate their source frame masses. %maintaining cWB's position as one of the leading searches for unexpected binary black hole (BBH) mergers. 
%we compare the sensitivity of different detector networks and discuss possible causes for the suboptimality. 
For simulated events distributed uniformly in a fiducial volume,  %corresponding to the redshift z = 2, 
we improve the detection efficiency with respect to the standard cWB search by approximately $20\%$ for both stellar-mass and intermediate mass black hole binary mergers, detected with a false-alarm rate less than $1\,\mathrm{yr}^{-1}$. We show the robustness of the ML-enhanced search for detection of generic BBH signals by reporting increased sensitivity to the spin-precessing and eccentric BBH events as compared to the standard cWB search. %, even when trained on simulated quasi-circular BBH events with aligned spins. 
Furthermore, we compare the improvement of the ML-enhanced cWB search for different detector networks.

\end{abstract}

\date[\relax]{Dated: \today }

\maketitle

%%%%%%%%%%%%%%%%%%%%%%%%%%%%%%%%%%%%%%%%%%%%%%%
\section{Introduction}
The Advanced LIGO~\cite{TheLIGOScientific:2014jea} and Advanced Virgo~\cite{TheVirgo:2014hva} network of detectors detected 11 GW candidates during the first two observing runs (O1 and O2)~\cite{GWTC1}, 44 GW candidates in the first half of the third observing run (O3a)~\cite{GWTC2, GWTC2.1}, and 35 GW candidates during the second half of the third observing run (O3b)~\cite{GWTC3}. These detections were identified by the different search pipelines employed by the LIGO-Virgo-KAGRA (LVK) collaboration, including GstLAL~\cite{GstLAL1, GstLAL2}, PyCBC~\cite{PyCBC1, PyCBC2}, MBTA~\cite{MBTA1}, and coherent WaveBurst (cWB)~\cite{Klimenko:2008fu, Klimenko:2016}.
%These detections were made possible by the two primary search types in LIGO: template based search and model independent search. While the template based searches like PyCBC and GstLAL, are exceptional in recovering GW signals from the strain data, they are limited to the waveforms that are well modelled and exist in the template bank.   

CWB is a search algorithm that looks for excess power in the time-frequency domain to identify GW signals in the LIGO-Virgo strain data~\cite{Klimenko:2008fu, Klimenko:2016, SoftwareX, Zenodo, cWB_homepage}.
Unlike matched-filter based pipelines, %other analysis pipelines that search for particular GW sources, 
cWB does not explicitly use signal waveform models, making it a model independent search.
Thus, the cWB pipeline is expected to play an integral role in the detection of poorly modeled or unexpected GW sources.
Historically, cWB was crucial for the discovery of the first binary black hole (BBH) merger GW150914~\cite{GW150914}, which initiated the era of GW astronomy. More recently, cWB contributed in the identification of higher multipoles for GW190814~\cite{GW190814}, an event associated with the coalescence of a binary system with the most unequal mass ratio yet measured with gravitational waves. cWB was also instrumental to the first direct detection of an intermediate mass black hole (IMBH) GW190521~\cite{GW190521.1-Discovery, cwb_GW190521}, which is the most massive and distant black hole merger observed via GWs. Overall, the cWB pipeline has contributed to the detection of 32 BBH events in the O1, O2, and O3 observing runs~\cite{GWTC1, GWTC2, GWTC3}.

The cWB algorithm has been recently used in combination with machine learning (ML) algorithms for various studies~\cite{Vinciguerra_2017, Cavagli__2020, Gayathri_2020, OBrien_2021}. In this paper, the standard cWB pipeline sensitivity to BBH mergers is enhanced by using the ML method as described in Ref.~\cite{MLPaper}. 
%, where the reanalysis of O1-O2 data resulted in the higher detection significance of the GW events.
We reanalyze the publicly available strain data from O3~\cite{GWOSC} using the ML-enhanced cWB pipeline for the 2-fold detector network consisting of LIGO Hanford and LIGO Livingston (HL). We optimize the ML-enhanced cWB search for the 3-fold detector network consisting of LIGO Hanford, LIGO Livingston, and Virgo (HLV), where we analyze limited data from O3 to compare the performance with the 2-fold detector network and discuss possible causes for sub-optimality of the 3-fold detector network.  %different detector networks. Using the ML-enhanced cWB search, we analyze limited data from the three detector network: LIGO Hanford, LIGO Livingston, and Virgo (HLV), to compare the performance with the two detector network and discuss possible causes of suboptimality.

The paper is organized as follows. In Section~\ref{sec:method}, we introduce the ML-enhanced cWB search pipeline and discuss the updates to the method.
Section~\ref{subsec:data} describes the data used to train and test our ML algorithm.
%In Section~\ref{sec:results}, 
We present the results of the O3 data reanalysis with the ML-enhanced cWB in Section~\ref{sec:results}. Section~\ref{subsec:openbox} reports the updated significance of BBH events detected during the O3 run. Section~\ref{subsec:marginal} discusses the marginal events found in this reanalysis, followed by the comparison of search sensitivity of the ML-enhanced cWB search against the sensitivity of the standard cWB search in Section~\ref{subsec:sensitivity}. The tests of robustness of the ML method are given in Section~\ref{subsec:Robustness} and Section~\ref{subsec:HLvsHLV} compares the performance of the ML method on different network configurations.
Finally, in Section~\ref{sec:conclusions}, we state the conclusions of the ML-enhanced cWB search for binary mergers on O3.

\section{Method}
\label{sec:method}

\subsection{Coherent WaveBurst}
The cWB pipeline searches for transient GW signals by identifying events with excess power in the time-frequency domain. cWB reconstructs coherent events in multiple detectors with minimal assumptions on the signal model~\cite{Klimenko2008, Klimenko:2016}.
The time-frequency domain data is built from the detector strain data by using the Wilson Daubechies Meyer (WDM) wavelet transform~\cite{Necula:2012zz}. 
%, where the data is normalized by the root-mean-squared  of the detector noise.
The algorithm identifies events by clustering the nearby WDM wavelets with excess power above the average fluctuations of the detector noise.
The pipeline generates an event for each selected cluster and reconstructs the signal waveform using the constrained maximum likelihood method~\cite{Klimenko:2016}.
The cWB pipeline estimates various summary statistics for each event, describing the time-frequency structure, signal strength, and coherence across the detector network.
%The main detection statistic for the cWB generic GW search is the signal-to-noise ratio (SNR)  defined for the LIGO detector network as:
%\begin{equation} \label{eq:1}
%\eta_\mathrm{0} = \sqrt{\frac{E_\mathrm{c}}{2\,\text{max}\left(1,\chi^2\right)}} \,. %\end{equation}
%Here, $E_\mathrm{c}$ denotes the coherent energy estimated by cross-correlating the reconstructed signal waveforms across different detectors, and $\chi^2 = E_\mathrm{n} / N_{\mathrm{df}}$ where $E_\mathrm{n}$ is the estimated residual noise energy and $N_{\mathrm{df}}$ is the the number of independent wavelet amplitudes describing the event. The $\chi^2$ correction in Equation~\ref{eq:1}, which is close to unity for genuine GW events, reduces the non-Gaussian noise contribution.
The standard cWB detection statistic used to identify the GW events has been updated with respect to previous searches as given below:
\begin{equation}\label{eqn2}
\eta_\mathrm{0} = \sqrt{\frac{E_\mathrm{c}}{1+\chi^2(\text{max}(1, \chi^2)-1)}}\, \\,
% \begin{cases}
% \mathrm{where,\, }\chi^2 = E_\mathrm{n} / N_{\mathrm{df}}
% \end{cases}
\end{equation}
%where $W_{\mathrm{M}}$ is the chirp mass penalty factor and is given by $W_{\mathrm{M}} = F_{\mathrm{M}} \, \sqrt{e_{\mathrm{M}}}$. Here, 
%where $F_{\mathrm{M}}$ is the event energy fraction, $e_{M}$ is the event ellipticity defined in Ref.~\cite{Tiwari2015_chirp} and $\eta_\mathrm{0}$ is the signal-to-noise ratio (SNR). Both the parameters  $F_{\mathrm{M}}$ and $e_{M}$ are close to unity for BBH events and penalize events which have a time-frequency evolution that is significantly different from the chirping BBH signal. 
Here, $\chi^2$ is given by $\chi^2 = E_\mathrm{n} / N_{\mathrm{df}}$ where $E_\mathrm{n}$ is the estimated residual noise energy, and $N_{\mathrm{df}}$ is the number of independent wavelet amplitudes describing the event. The $\chi^2$ factor is close to unity for genuine GW events and penalizes the non-Gaussian noise contribution. The coherent energy $E_\mathrm{c}$, is estimated by cross-correlating the reconstructed signal waveforms across different detectors.
In the standard cWB analysis, the significance of GW events is increased by removing the excess background with \textit{a priori} defined veto conditions on a set of summary statistics generated for each event. Although this method works well, it may entirely remove the borderline GW events which do not pass the veto thresholds. The veto conditions need to be manually tuned for each observing run and detector network. Moreover, designing these veto conditions in the multi-dimensional space of the summary statistics is challenging. Using the ML method, allows us to automate and enhance the separation of signal and noise events while simultaneously increasing the sensitivity of the search~\cite{MLPaper}. 
%GW detector data is hindered by noise artifacts known as glitches, and consequently, some noise events are reconstructed by the pipeline and leak into the analysis. In the standard cWB analysis, we apply a series of vetoes to target and remove these glitches. 
%This approach, henceforth known as the \textit{veto method}, improves the significance of candidate GW events by reducing excess background. 
%The veto method consists of applying a priori defined veto thresholds on a set of the cWB summary statistics. This procedure discretely classifies generated events into one of the two categories: signal-like events and noise-like events. Events that fall into the noise-like category are removed from the analysis.
%This process could inevitably result in discarding borderline GW events which do not pass the veto thresholds and at the same time makes the pipeline vulnerable to the high SNR glitches which do pass the vetoes. Designing vetoes in the multidimensional space of the summary statistics is challenging, and furthermore, requires re-tuning of the veto thresholds for each detector network configuration and each observing run.
In the ML-enhanced cWB pipeline, we define the reduced detection statistic given by:
\begin{equation}\label{eq:2}
\eta_\mathrm{r} = \eta_\mathrm{0}\cdot W_{\mathrm{XGB}}, 
\end{equation}
where $W_{\mathrm{XGB}}$ is the penalty factor provided by the ML algorithm~\cite{MLPaper}. 

In the standard cWB search, we employ two different search configurations to improve the search sensitivity: the BBH configuration targeting stellar-mass BBH mergers ($M_{\mathrm{tot}} \lesssim 100\, M_\odot$) and the IMBH configuration targeting intermediate mass  black hole (IMBH) binary mergers ($M_{\mathrm{tot}} \gtrsim 100\, M_\odot$). The corresponding GW signals observed in the LIGO frequency band are quite different for the two systems. A GW signal originating from the stellar-mass BBH merger and observable in the LIGO bandwidth usually exhibits the entire inspiral-merger-ringdown waveform. In contrast, GW signals from IMBH binary mergers are short in duration and contain mostly the merger-ringdown waveform, with the inspiral signal buried inside the low-frequency seismic noise. A specialized search for IMBH mergers helps to constrain the PISN mass gap~\cite{POP_GWTC2} and possible formation channels~\cite{21gImplications} for massive binaries.
%Studying IMBH mergers is astrophysically enriching, as it helps us to understand the IMBH population and estimate the corresponding merger rates. It also enables studies to constrain the PISN mass gap and possible formation channels.
%As a result, we utilize two different cWB search configurations for these systems: the BBH configuration, which targets stellar-mass BBH mergers, and the IMBH configuration, which targets IMBH binary mergers. 

The estimated central frequency $f_0$ of a GW signal is inversely proportional to the red-shifted total mass of the binary system.
As a result, IMBH binaries are expected to merge at lower frequencies than the stellar-mass BBH mergers. So, we select events with $60\, \mathrm{Hz} < f_0 < 300\, \mathrm{Hz}$ for the BBH search configuration and events with $f_0 < 200\, \mathrm{Hz}$ for the IMBH search configuration. Since the two search configurations overlap in the frequency bands, we combine them in accordance with the rules explained in Table 1 of Ref.~\cite{cwb_GW190521}.
%consider trial factors for the calculation of the detection significance. The trial factors are calculated based on the peak frequencies of the reconstructed events, in accordance with the rules explained in Table 1 of Ref.~\cite{cwb_GW190521}.  

\subsection{Data}\label{subsec:data}
We analyze publicly available strain data from Advanced LIGO and Advanced Virgo's third observational run~\cite{GWOSC}. In order to train and test a supervised ML algorithm, we generate two types of data: noise events (background) and signal events (simulations). We accumulate the background data by time-shifting the data from one detector with respect to other detectors in the network. The time-shifted data is analyzed with the cWB pipeline to %acquire
estimate the characteristics of the background events. To generate the simulation data, we inject simulated GW signals into the detector data and recover these signals using the cWB pipeline. In this study, we use four sets of simulations: (i) a %quasi-circular 
spin-aligned stellar-mass BBH set with quasi-circular orbit approximation, (ii) a numerical relativity (NR) %a quasi-circular 
IMBH binary set, (iii) an eccentric BBH set, and (iv) a precessing BBH set with quasi-circular orbit approximation. The binary orientation parameters (sky location, inclination angle) for every simulated waveform are randomly drawn from uniform distributions in all four cases. The redshift $z$ is drawn from a uniform distribution in co-moving volume, assuming Planck 2015 cosmology~\cite{Planck2015}. Simulation sets (i) and (ii) are used to train and test the ML algorithm, whereas the remaining two sets are only used to test the robustness of the ML implementation. 

%BPL, PLP
In the simulation set (i), the component black hole masses are drawn from models representing the stellar-mass BBH population~\cite{Broken_power_law, Power_law_peak}. The source frame total mass for these simulations ranges from approximately $5\,M_\odot$ to $100 \, M_\odot$, and the mass ratios range from 1 to 1/4. We use the SEOBNRv4~\cite{Boh2017} waveform approximant to generate the simulations. The spins of the component black holes are drawn from a uniform, aligned spin distribution. The binary source distance is randomly drawn, assuming the uniform density in co-moving volume. %up to redshift z = 2.
%[Place holder for BBH set] To simulate the stellar-mass BBH set, we use the SEOBNRv3~\cite{Babak:2016tgq} and SEOBNRv4~\cite{Boh2017} waveform approximants.
%These waveform approximants include only the dominant ($\ell = 2, m = 2$) harmonic mode.
%The source frame total mass for these simulations ranges from approximately $5\,M_\odot$ to $100 \, M_\odot$, and the mass ratio $q = m_2/m_1$ ranges from approximately 1/4 to 1.
%Component black hole spins are aligned with the orbital plane.
%-some spins drawn from aligned distribution, some spins drawn from isotropic distribution

%NRALL SIM
For simulation set (ii), we use NR waveforms, including higher-order harmonics, as the representative models for IMBH binary set. We consider 17 mass bins, as described in Ref.~\cite{Abbott:2019ovz}, ranging in the source frame total mass from $120\, M_\odot$ to $800 \, M_\odot$, and with mass ratios ranging from 1 to 1/10.

In simulation set (iii), we also use NR waveforms for the high mass, eccentric BBH set~\cite{Healy2017, gayathri2020gw190521}. We consider 28 mass bins ranging in total mass from $100 \, M_\odot$ to $250 \, M_\odot$, with mass ratio equal to 1, and eccentricities ranging from 0.19 to 0.96.

For the simulation set (iv) involving precessing stellar-mass BBH, we use the SEOBNRv4PHM~\cite{SEOBNRV4PHM} waveform approximant, which includes precession and the higher-order harmonic modes.
The source frame total mass ranges from $4\, M_\odot$ to $200 \, M_\odot$, with mass ratios ranging from 1 to 1/20.
The spins of the component black holes are isotropically distributed.

The amount of simulation and accumulated background data used for training and testing the ML algorithm is described in Tab.~\ref{tab:1}.

\begin{table}[bht]
    \centering
    \setlength{\tabcolsep}{3pt}
    
    \begin{tabular}{lclcl}
        \hline
        \hline
        O3 Dataset & \multicolumn{2}{c}{Background [yr]} & \multicolumn{2}{c}{Simulation [events]}\\
        & Training & Testing & Training & Testing\\
        \hline
        \hline
        BBH & $2000$ & $28527$ & $20077$ & $110861$\\
        IMBH & $2000$ & $28535$ & $19801$ & $192626$\\
        %BBH & $1100$ & $19093$ & $10829$ & $49416$\\
        %IMBH & $1100$ & $19101$ & $5484$ & $103639$\\
        
        Eccentric BBH & $-$ & $19101$ & $-$ & $974576$\\
        Precessing BBH & $-$ & $19093$ & $-$ & $17483$\\
        
        Network Study  & $550$ & \multicolumn{1}{c}{$639$} & $1998$ & $1992$\\
        %BBH  & \multirow{2}{*}{$550$} & \multirow{2}{*}{$639$} & \multirow{2}{*}{$1998$} & \multirow{2}{*}{$2400$}\\
        %(HLV, limited data) & & & & \\
    
        \hline
        \hline
    \end{tabular}
    \caption{Amount of accumulated background data (given in years) and the number of simulated events used for training and testing. We consider the third observing run (O3) for the two detector network (HL), search configurations (BBH, IMBH), and additional simulation cases to test the model robustness (Eccentric BBH, Precessing BBH). For the network study, we also consider limited BBH data (1/33rd of the O3 run) for the three detector network (HLV).
    }
    \label{tab:1}
\end{table}

\subsection{XGBoost}
We describe the procedure followed to train the ML algorithm and discuss updates on the method described in Ref.~\cite{MLPaper}. Here, we use an ensemble learning, boosted decision-tree based ML algorithm called XGBoost~\cite{XGBoost}. We start by selecting the subset of the cWB summary statistics used as the list of input features for training and testing the XGBoost algorithm. The data is split into training and testing as defined in Tab.~\ref{tab:1}, and we train two different models over the entire O3 run for the HL network: BBH search configuration and IMBH search configuration. For the detector network comparison discussed in section~\ref{subsec:HLvsHLV}, we train a BBH search configuration model for the HLV network with half of the limited data which is 1/33rd of the O3 run (GPS: 1241011102.0 to GPS: 1242485126), given in Tab.~\ref{tab:3}. We test the trained models on the remaining background and simulation data for each search type, apply these models to the GW candidate events in O3, and present the results in section~\ref{sec:results}.

The ML training and testing method utilized in this paper is consistent with the one presented in Ref.~\cite{MLPaper} except for the following changes: change in the subset of summary statistics used as the input features for XGBoost, the values of the XGBoost hyper-parameters used for training the models, the introduction of the standardized sample weight. These updates to the existing method are discussed in Appendix~\ref{XGB}.

\section{Results}
\label{sec:results}
In this section, we discuss the results of the ML-enhanced cWB search on O3 data. We first report the detections in subsection~\ref{subsec:openbox}, followed by the discussion of marginal events in subsection~\ref{subsec:marginal}. In Section~\ref{subsec:sensitivity}, the improvement in the search sensitivity is discussed. We include tests of robustness to check the ML implementation in subsection~\ref{subsec:Robustness}. Lastly, we compare the performance of searches with different detector networks in subsection~\ref{subsec:HLvsHLV}.%, by analyzing limited data (GPS: 1241011102.0 to GPS: 1242485126) of the BBH search configuration for the HLV detector network. 

\subsection{Search results}\label{subsec:openbox}

We train two separate models over the entire O3 run, one for the BBH search configuration and the other for the IMBH search configuration. The trained models are used to calculate the $W_{\mathrm{XGB}}$, which is required for the estimation of the reduced detection statistic $\eta_\mathrm{r}$ (defined in Equation~\ref{eq:2}) for all candidate events. 
%The ML-enhanced cWB search uses the reduced detection statistic $\eta_\mathrm{r}$, whereas the standard cWB uses $\eta_\mathrm{1}$ as the detection statistic for detections. 
The detection statistic provides the ranking of detected events and allows the calculation of the false-alarm rate (FAR), which is used to assign the significance of a candidate event. The FAR value is estimated by counting the number of background events with an equal or higher value of the detection statistic than for the given candidate event, divided by the total available background time.

\begin{table*}[bht]
\centering
    \setlength{\tabcolsep}{4pt}
 %\begin{tabular}{p{2.0cm}cc}
  \begin{tabular}{lcccc}
        \hline
        \hline
         Event & \multicolumn{1}{c}{Standard cWB} & \multicolumn{1}{c}{ ML-enhanced cWB} & SNR & \pastro \\
        %Event & \multicolumn{1}{c}{($\eta_\mathrm{1}$ + vetoes)} & \multicolumn{1}{c}{($\eta_{r}$)} & SNR \\
         & \multicolumn{1}{c}{FAR [yr$^{-1}$]} & \multicolumn{1}{c}{FAR [yr$^{-1}$]} & & \\[0.5ex]
        \hline
        \hline 
        
        GW190408\_181802 & $< 9.5 \times 10^{-4}$ & $< 1.0 \times 10^{-3}$ & $14.8$ & $0.999$ \\
        GW190412 & $< 9.5 \times 10^{-4}$ & $< 1.0 \times 10^{-3}$ & $19.7$ & $0.999$\\
        GW190421\_213856 & $\quad\, 3.0 \times 10^{-1}$ & $\quad\,1.8 \times 10^{-2}$ & $9.3$ & $0.997$\\
        GW190503\_185404 & $\quad\,1.8 \times 10^{-3}$ & $< 9.9 \times 10^{-4}$ & $11.5$ & $0.999$\\
        GW190512\_180714 & $\quad\,3.0 \times 10^{-1}$ & $\quad\,1.8 \times 10^{-1}$ & $10.7$ & $0.941$\\
        GW190513\_205428 & \multicolumn{1}{c}{$...$} & $\quad\,1.0 \times 10^{+0}$ & $11.5$ & $0.703$\\
        GW190517\_055101 & $\quad\,6.5 \times 10^{-3}$ & $\quad\,6.2 \times 10^{-4}$ & $10.7$ & $0.999$\\
        GW190519\_153544 & $\quad\,3.1 \times 10^{-4}$ & $< 1.0 \times 10^{-4}$ & $14.0$ & $1.000$\\
        GW190521 & $\quad\,2.0 \times 10^{-4}$ & $< 1.0 \times 10^{-4}$ & $14.4$ & $1.000$\\
        GW190521\_074359 & $< 1.0 \times 10^{-4}$ & $< 1.0 \times 10^{-4}$ & $24.7$ & $0.999$\\

        GW190602\_175927 & $\quad\,1.5 \times 10^{-2}$ & $< 8.8 \times 10^{-4}$ & $11.1$ & $0.999$\\
        GW190701\_203306 & $\quad\,5.5 \times 10^{-1}$ & $\quad\,1.1 \times 10^{-2}$ & $10.2$ & $0.997$\\
        GW190706\_222641 & $< 1.0 \times 10^{-3}$ & $< 1.1 \times 10^{-3}$ & $12.7$ & $0.999$\\
        GW190707\_093326 & \multicolumn{1}{c}{$...$} & $\quad\,1.1 \times 10^{-1}$ & $11.2$ & $0.976$\\
        GW190727\_060333 & $\quad\,8.8 \times 10^{-2}$ & $\quad\,3.4 \times 10^{-3}$ & $11.4$ & $0.998$\\
        GW190728\_064510 & \multicolumn{1}{c}{$...$} & $\quad\,2.6 \times 10^{-2}$ & $10.5$ & $0.993$\\
        GW190828\_063405 & $< 9.6 \times 10^{-4}$ & $< 1.1 \times 10^{-3}$ & $16.6$ & $0.999$\\
        GW190915\_235702 & $< 1.0 \times 10^{-3}$ & $< 1.1 \times 10^{-3}$ & $12.3$ & $0.999$\\
        GW190929\_012149 & \multicolumn{1}{c}{$...$} & $\quad\,7.7 \times 10^{-1}$ & $9.2$ & $0.542$\\
        %\hline
        %GPS: 1251654568.58 & $...$ & $4.3 \times 10^{-1}$\\
        %\multicolumn{3}{l}{\textcolor{red}{Trigger removed by sub60Hz clean data}} \\
        GW191109\_010717 & $< 1.1 \times 10^{-3}$ & $< 1.2 \times 10^{-3}$ & $15.6$ & $0.999$\\
        GW191127\_050227 & \multicolumn{1}{c}{$...$} & $\quad\,2.1 \times 10^{-1}$ & $8.6$ & $0.957$\\
        GW191204\_171525 & $< 8.7 \times 10^{-4}$ & $< 9.6 \times 10^{-4}$ & $17.1$ & $0.999$\\
        GW191215\_223052 & $\quad\,1.2 \times 10^{-1}$ & $< 9.6 \times 10^{-4}$ & $9.8$ & $0.998$\\
        GW191222\_033537 & $< 8.9 \times 10^{-4}$ & $< 9.8 \times 10^{-4}$ & $11.1$ & $0.999$\\
        GW191230\_180458 & $\quad\,5.0 \times 10^{-2}$ & $\quad\,2.0 \times 10^{-3}$ & $10.3$ & $0.998$\\
        GW200128\_022011 & $\quad\,1.3 \times 10^{+0}$ & $\quad\,2.2 \times 10^{-2}$ & $8.8$ & $0.985$\\
        GW200219\_094415 & $\quad\,7.7 \times 10^{-1}$ & $\quad\,2.7 \times 10^{-1}$ & $10.4$ & $0.887$\\
        GW200224\_222234 & $< 8.8 \times 10^{-4}$ & $< 9.7 \times 10^{-4}$ & $18.8$ & $0.999$\\
        GW200225\_060421 & $< 8.8 \times 10^{-4}$ & $< 9.7 \times 10^{-4}$ & $13.1$ & $0.999$\\
        GW200311\_115853 & $< 8.2 \times 10^{-4}$ & $< 8.9 \times 10^{-4}$ & $16.2$ & $0.999$\\
        
        \hline
    \end{tabular}
    \caption{O3 ML-enhanced cWB search results compared with the standard cWB results. We report all detections with FAR $\leq1\, \mathrm{yr}^{-1}$ and present in the catalog papers~\cite{GWTC2, GWTC3}. The estimated significance is limited by the accumulated background data and is indicated with a '$<$' entry. For the ML-enhanced cWB search, the background data used for training the ML model has been removed from the total available accumulated background data. We estimate the \pastro and report it along with the SNR reconstructed by the cWB pipeline for the GW events.}
\label{tab:2}
\end{table*}

Tab.~\ref{tab:2} reports the BBH candidates identified by the ML-enhanced cWB search in the O3 observing run. We identify with equivalent or higher significance all 15+10 (O3a+O3b) BBH candidates previously reported by the standard cWB search~\cite{GWTC2, GWTC3}. Moreover, the ML-enhanced cWB search detected GW190513, GW190707, GW190728, GW190929, and GW191127 with FAR of $<1\,\mathrm{yr}^{-1}$, which were previously vetoed in the standard cWB search, but  detected by the other search pipelines~\cite{GWTC2, GWTC3}. Three cWB-only marginal candidate events, not listed in the LVK catalogs, were also found by the ML-enhanced cWB search over O3 and are discussed in subsection~\ref{subsec:marginal}. For all the GW and marginal candidate events in O3, we compute the corresponding probabilities of astrophysical origin (\pastro) assuming that the terrestrial and astrophysical events occur as independent Poisson processes~\cite{pastro, GWTC3}.
%To comment on the sensitivity of the ML-enhanced cWB search, 

\subsection{Marginal candidate events}\label{subsec:marginal}
\begin{figure*}[th]
    \centering
    \includegraphics[width=\linewidth]{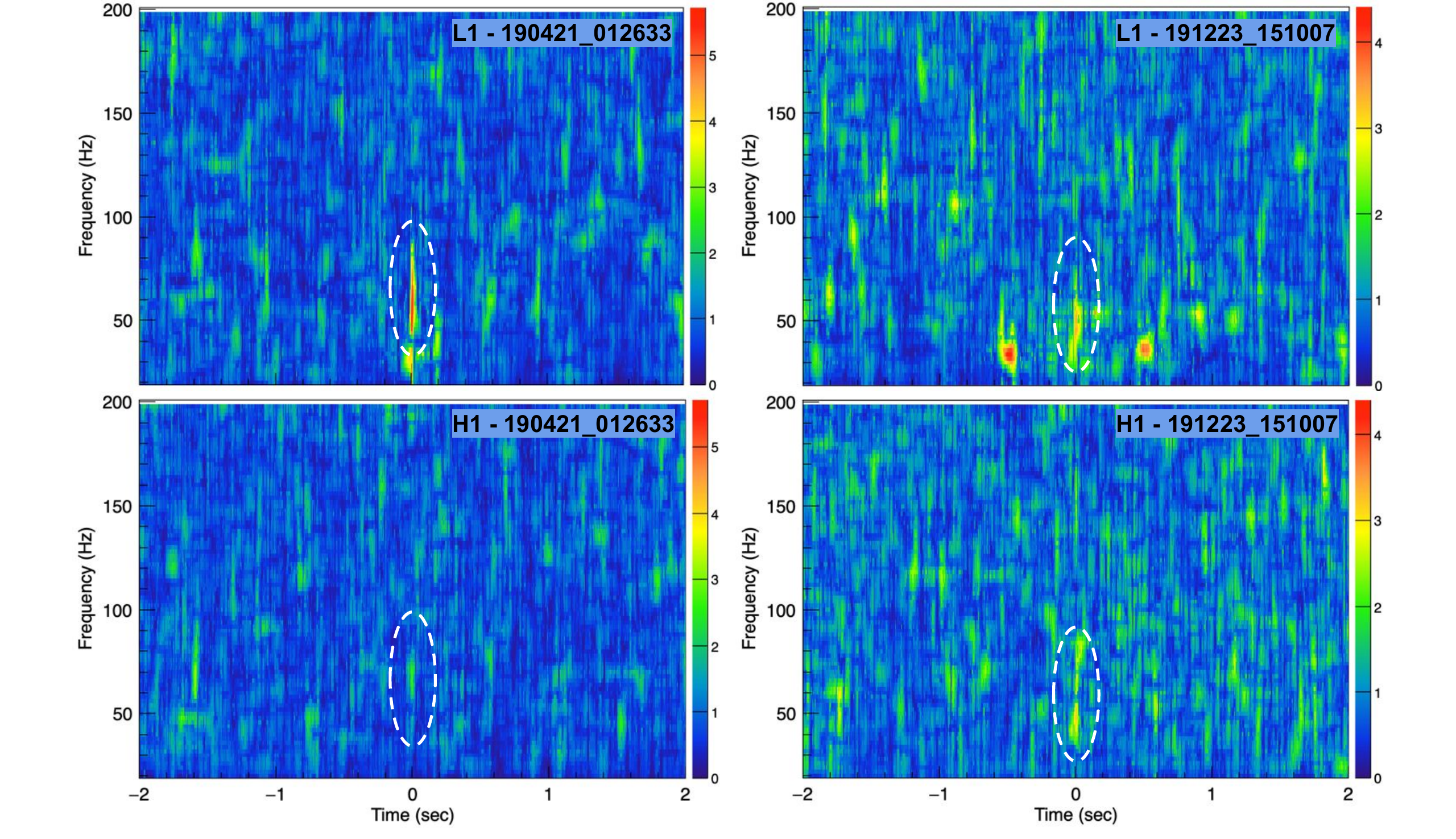}
    \caption{Time-frequency wavescan spectrograms~\cite{wavescan} of 190421\_012633 (LEFT) and 191223\_151007 (RIGHT), in the L1: Livingston (TOP) and H1: Hanford (BOTTOM) detectors. The marginal events occur at Time $=$ 0 seconds in the plots and are highlighted by white dashed circles.}
    \label{fig:wavescan_190421}
\end{figure*}

In GWTC-3, the cWB search had reported three marginal events, namely 190804\_083543, 190930\_234652, and 200214\_224526~\cite{GWTC3}. These events were identified by the standard cWB pipeline and are consistent with noise artifacts found around their GPS times. The ML-enhanced cWB search also identifies these events but assigns them a lower significance: 190804\_083543 with a FAR of 28 $\mathrm{yr}^{-1}$, 190930\_234652 with a FAR of 8 $\mathrm{yr}^{-1}$, and 200214\_224526 with a FAR of 17 $\mathrm{yr}^{-1}$. The significance of these events is below the analysis threshold of FAR $<$ 1 $\mathrm{yr}^{-1}$, and hence, they are not included in this paper. Instead, in Tab.~\ref{tab:3}, we report the GW candidate events that are not present in the GW Transient Catalog papers but were detected by the ML-enhanced cWB  search with FAR $<1\, \mathrm{yr}^{-1}$. The lowest-ranked event 190904\_174910 is detected with a FAR of $0.85\, \mathrm{yr}^{-1}$ and reconstructed SNR of $8.5$.
%The \pastro value is $0.573$, which does not favor the astrophysical origin for this event.
The event 190904\_174910 is consistent with the instrumental noise produced around the 60Hz line by the coupling of the low-frequency seismic noise~\cite{Driggers2019, Davis:2018yrz, Vajente2019}. %and this event is consequently removed from further analysis. %upconversion seismic noise ... reanalysed with sub60Hz 

\begin{table*}[bht]
\centering
    \setlength{\tabcolsep}{4pt}
 %\begin{tabular}{p{2.0cm}cc}
  \begin{tabular}{lrccccccc}
        \hline
        \hline
         Event & \multicolumn{1}{c}{Standard cWB} & \multicolumn{1}{c}{ ML-enhanced cWB} & SNR & \pastro & $M_{\mathrm{cWB}}$ & $M$ & $m_{\mathrm{1}}$ & $m_{\mathrm{2}}$\\
        %Event & \multicolumn{1}{c}{($\eta_\mathrm{1}$ + vetoes)} & \multicolumn{1}{c}{($\eta_{r}$)} & SNR & $M_{\mathrm{cWB}}$ & Remarks\\
         & \multicolumn{1}{c}{FAR [yr$^{-1}$]} & \multicolumn{1}{c}{FAR [yr$^{-1}$]} & & & $ [M_{\odot}]$ & $ [M_{\odot}]$ & $ [M_{\odot}]$ &$ [M_{\odot}]$ \\ [0.5ex]
        \hline
        \hline
        
        190421\_012633 & $7.1 \times 10^{+0}$ & $4.4 \times 10^{-1}$ & $9.6$ & $0.833$ & 160 & $155^{+65}_{-32}$ & $111^{+61}_{-38}$ & $44^{+25}_{-20}$\\
        %190421\_012633 & $7.1 \times 10^{+0}$ & $4.4 \times 10^{-1}$ & $9.6$ & $0.833$ & 160 & $154.7^{+65.3}_{-31.5}$ & $111.2^{+61.4}_{-38.2}$ & $44.1^{+24.6}_{-19.8}$\\
        190904\_174910* & \multicolumn{1}{c}{---} & $8.5 \times 10^{-1}$ & $8.1$ & $0.573$ & --- & --- & --- & ---\\
        191223\_151007 & \multicolumn{1}{c}{---} & $4.3 \times 10^{-1}$ & $7.5$ & $0.742$ & 141 & $206^{+55}_{-39}$ & $153^{+43}_{-38}$ & $54^{+27}_{-19}$\\
        %191223\_151007 & \multicolumn{1}{c}{$...$} & $4.3 \times 10^{-1}$ & $7.5$ & $0.742$ & 141 & $206.0^{+55.1}_{-39.4}$ & $152.8^{+42.5}_{-37.5}$ & $53.8^{+26.7}_{-18.7}$\\

        \hline
    \end{tabular}
    \caption{Marginal cWB-only events were found in O3 with the ML-enhanced cWB search compared with the standard cWB analysis. We report all detections with FAR $<1\, \mathrm{yr}^{-1}$ not present in the catalog papers. We estimate the \pastro, cWB source frame total mass ($M_{\mathrm{cWB}}$) and report it along with the SNR reconstructed by the cWB pipeline for the marginal events. We report the median and 90\% symmetric credible intervals for the one-dimensional marginal posterior distributions of the source total mass $M$, and the source component masses: $m_{\mathrm{1}}$ and $m_{\mathrm{2}}$. *190904\_174910 is likely to be produced by the up-conversion of seismic noise and the 60Hz power line. }
\label{tab:3}
\end{table*}

The other two cWB-only events, 190421\_012633 and 191223\_151007 are marginal events detected  with the FAR of $< 0.44\, \mathrm{yr}^{-1}$ and $< 0.43\, \mathrm{yr}^{-1}$ and reconstructed SNR of 9.6 and 7.5, respectively. %While these two events do not have sufficient significance to claim confident detection, we have not identified any instrumental noise origin for these events, making them marginal candidate events. 
We follow the XGBoost parameter estimation technique described in Ref.~\cite{OBrien_2021} and use their trained model to estimate the source frame total mass ($M_{\mathrm{cWB}}$). The highest-ranked event 191223\_151007 is estimated to have the $M_{\mathrm{cWB}}$ of $141\, M_{\odot}$ with the \pastro value of $0.742$. The standard cWB analysis had previously removed this event as it had failed the \textit{a priori} defined veto thresholds on the event summary statistics. 
%This event was removed by the standard cWB analysis as it failed the \textit{a priori} defined vetoes on \texttt{norm} ($n_f$) and \texttt{TFcut} summary statistics. 
The event 190421\_012633 was reconstructed in the standard cWB analysis with a FAR $7.1\, \mathrm{yr}^{-1}$, and it was promoted by the ML-enhanced cWB search to a marginal event with a lower FAR. This event is estimated to have the $M_{\mathrm{cWB}}$ of $160\, M_{\odot}$ with the \pastro value of $0.833$. 

\begin{figure}[h!]
    \centering
    \includegraphics[width=\linewidth]{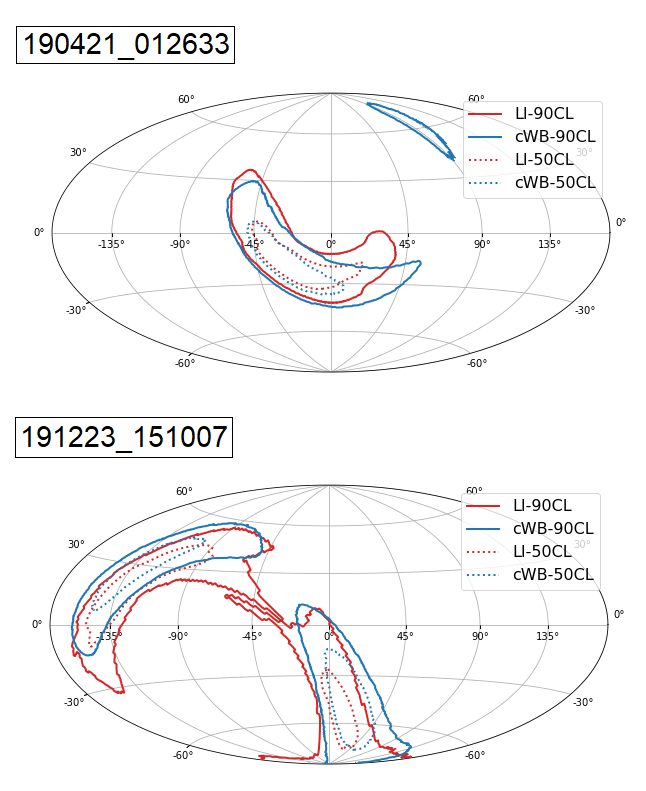}
    \caption{The 90\% and 50\% credible sky localization area estimated by cWB (Blue) vs the LAL Inference posterior distributions (Red), plotted for the marginal events 190421\_012633 and 191223\_151007.}
    \label{fig:skyloc}
\end{figure}

% \begin{figure}[h!]
%     \centering
%     \includegraphics[width=\linewidth]{o3b191223_LH.png}
%     \caption{Time-frequency wavescan of 191223\_151007 in the L1 (Livingston) and H1 (Hanford) detectors.}
%     \label{fig:wavescan_191223}
% \end{figure}

In order to extract the source properties of the two marginal events, we perform the coherent Bayesian parameter estimation \cite{Veitch2015, Lange2017, Wysocki2019}. In this analysis, we model the GW signal as represented by a precessing quasi-circular waveform IMRPhenomXPHM ~\cite{Pratten:2020ceb, Pratten:2020fqn, Garcia-Quiros:2020qpx, Garcia-Quiros:2020qlt}.  %In order to extract the astrophysical source prope of the two marginal events, we do followed up with a coherent Bayesian parameter estimation analysiswe draw discrete samples using the IMRPhenomXPHM~\cite{Pratten:2020ceb, Pratten:2020fqn, Garcia-Quiros:2020qpx, Garcia-Quiros:2020qlt} waveform and produce the posterior probability distributions for the source properties using Bayesian inference techniques (need to check with Gayathri). 
We report the median and 90\% symmetric credible intervals for the one-dimensional marginal posterior distributions of the source total mass $M$ and the source component masses $m_{\mathrm{1}}$ and $m_{\mathrm{2}}$ in Tab.~\ref{tab:3}. The source frame total masses for both events are in agreement with the cWB estimated total masses. The secondary mass components fall within the stellar mass BBH range for both events, whereas the primary mass components fall in the IMBH range with the source masses greater than $100\, M_{\odot}$. 
%. The primary component source mass of 191223\_151007 event is   190421\_012633 is $152.76^{+42.5}_{-37.46}\, M_{\odot}$ whereas 190421\_012633 has $111.22^{+61.35}_{-38.17}\, M_{\odot}$ 
%making them more like the IMBH event GW190521 detected by LIGO~\cite{GW190521.1-Discovery, cwb_GW190521}. 
 
The sky localization area for the 190421\_012633 and 191223\_151007 marginal events are shown in Fig.~\ref{fig:skyloc}.
Here, the 90\% credible sky area estimated by cWB is consistent with the sky area inferred from the Bayesian posterior distribution. 
Fig.~\ref{fig:wavescan_190421} displays the time-frequency wavescan spectrograms~\cite{wavescan} of 190421\_012633 and 191223\_151007. Although the low significance of these events is insufficient to claim confident detection, we have not identified them with any common instrumental noise origin.   

\subsection{Search sensitivity}\label{subsec:sensitivity}
\begin{figure}[h!]

    \centering
    \includegraphics[width=\linewidth]{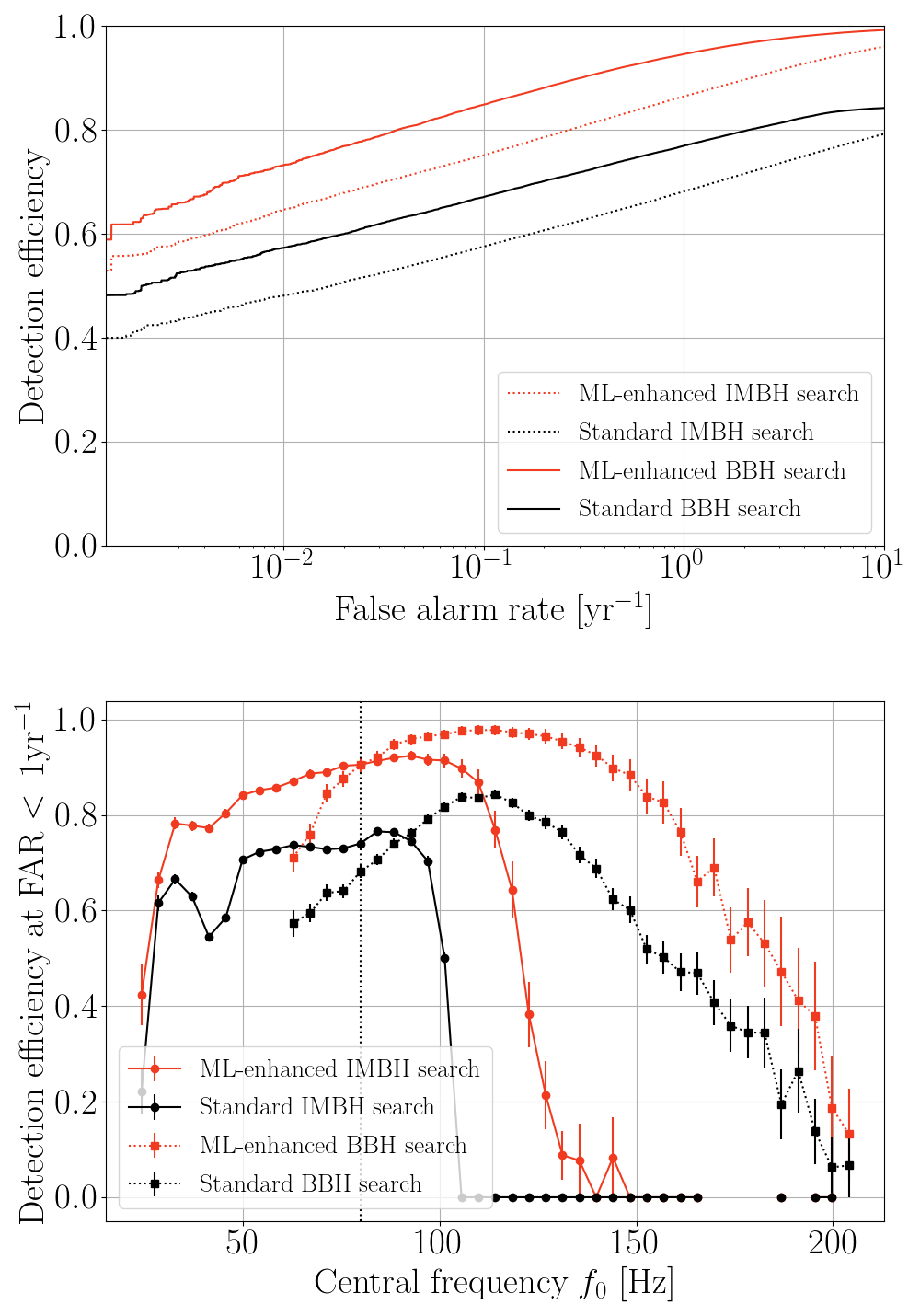}
    \caption{Top: Detection efficiency vs FAR for the O3 observing run. Bottom: Detection efficiency for events identified with FAR less than $1\,$yr$^{-1}$ as a function of the central frequency $f_0$ for O3. Solid lines correspond to the BBH configuration, while dotted lines represent the IMBH configuration. Red curves represent the sensitivity of the ML-enhanced cWB search, and blue curves represent the sensitivity of the standard cWB search with the veto method.}
    \label{fig:efreq}

\end{figure}

The detection efficiency is calculated by taking the number of detected simulated events with FAR equal to or less than a given threshold and dividing it by the total number of recovered simulation events. From the detection efficiency vs FAR in Fig.~\ref{fig:efreq}, we observe an 18\% improvement for both the BBH and IMBH configurations. For simulated events detected with FAR $<\,1\,\mathrm{yr}^{-1}$, in Fig.~\ref{fig:efreq} we plot the detection efficiency as a function of the central frequency $f_0$.
%The ML-enhanced cWB search shows improved sensitivity as compared to the standard cWB search for both search configurations. 
%The detection efficiency curves of the BBH and IMBH search configurations cross each other at $f_\mathrm{0} = 80$ Hz, and t
The overall ML-enhanced search is more sensitive than the standard search over the entire BBH mass range accessible by LIGO.

\begin{figure}[h!]

    \centering
    \includegraphics[width=\linewidth]{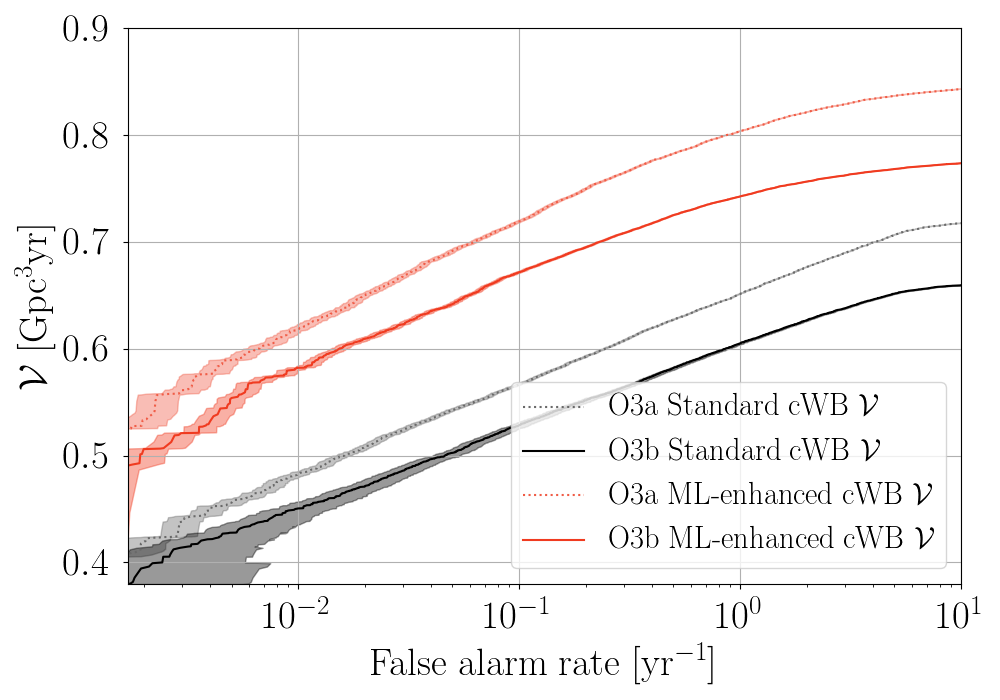}
    \caption{Sensitive hyper-volume $\mathcal{V}$ plotted against FAR for O3 cWB searches using the BBH \textsc{Power Law + Peak} simulations. Solid lines correspond to the O3b run, while dotted lines represent the O3a run. Red curves represent the sensitivity of the ML-enhanced cWB search, and black curves represent the sensitivity of the standard cWB search. The corresponding shaded regions denote the respective 1-$\sigma$ confidence intervals on the FAR.}
    \label{fig:volume}

\end{figure}
% \begin{figure}[h!]

%     \centering
%     \includegraphics[width=\linewidth]{Volume_plot_PLP_2.png}
%     \caption{The \% gain in the sensitive hyper-volume $\mathcal{V}$ by using the ML-enhanced cWB search with respect to the standard cWB search, plotted against FAR for the O3 cWB searches. The sensitive hyper-volume $\mathcal{V}$ is calculated by considering the BBH \textsc{Power Law + Peak} simulations. Solid line corresponds to the O3b run, while the dotted line represents the O3a run.}
%     \label{fig:volume1}

% \end{figure}

\begin{table}[bht]
    \centering

    \begin{tabular}{lcccccc}
        \hline
        \hline
         & \multicolumn{6}{c}{Sensitive hyper-volume $\mathcal{V}$ (Gpc$^3$yr)}\\ 
        Run & Std. & ML-enhanced & GstLAL & MBTA & PyCBC & PyCBC\\
        & cWB & cWB &  &  &  & BBH\\
        \hline
        \hline
        O3a & $0.676$ & $0.822$ & $1.22$ & $0.885$ & $0.914$ & $1.20$\\
        O3b & $0.627$ & $0.757$ & --- & --- & --- & ---\\
        \hline
        \hline
    \end{tabular}
    \caption{The effective hyper-volume $\mathcal{V}$ for the \textsc{Power Law + Peak} population model injections over the O3 run, estimated at a FAR of 2 $\mathrm{yr}^{-1}$. We compute the $\mathcal{V}$ for the standard and ML-enhanced cWB searches over the O3a and O3b data. We compare the O3a hyper-volumes for all the searches, including GstLAL, MBTA, PyCBC, and PyCBC BBH searches found in Ref.~\cite{GWTC2}. 
    }
    \label{tab:5}
\end{table}
We compare the sensitivity of the ML-enhanced cWB pipeline with the other search pipelines by computing the sensitive hyper-volume $\mathcal{V}$ in units of volume$\times$time~\cite{GWTC2}. For this study, we choose the \textsc{Power Law + Peak} model~\cite{GWTC2}, which is a fiducial BBH population model with a high posterior probability in the population analysis of GW Transient Catalog 2~\cite{POP_GWTC2}. Given a local merger rate $R(z=0)$ and the expected number of detections $N_{\mathrm{det}}$ at a given significance threshold, we estimate the effective hyper-volume $\mathcal{V}$ by following the relation $N_{\mathrm{det}}=\mathcal{V} R(z=0)$, where $N_{\mathrm{det}}$ is calculated by counting the number of injections detected by the search pipeline above a threshold at FAR 2 $\mathrm{yr}^{-1}$. The effective hyper-volume for each search for the given signal population is given in Tab.~\ref{tab:5}.

We note that the ML-enhanced cWB search improves the sensitive hyper-volume by 0.13 - 0.15 Gpc$^3$yr compared to the standard cWB search. In the O3a run, the ML-enhanced cWB sensitive hyper-volume is comparable to the MBTA search pipeline~\cite{GWTC2}. Fig.~\ref{fig:volume} shows the sensitive hyper-volume against FAR for the ML-enhanced cWB and the standard cWB search for the O3a and O3b runs. 
%Fig.~\ref{fig:volume1} shows the \% gain in the sensitive hyper-volume against FAR for the ML-enhanced cWB with respect to the standard cWB search, where we observe around $20-30\%$ gain in both the O3a and O3b runs for FAR $<$ 1 per year. 

\subsection{Test of Robustness}\label{subsec:Robustness}
The ML implementation is designed such that we maintain the model independent nature of the cWB search. The ML-enhanced search sensitivity is not limited to the simulations present in the training set~\cite{MLPaper}. In order to check this, we analyze the performance of the ML-enhanced search on simulated waveforms outside the training set. 
We train the ML model on quasi-circular binaries and test the ML-enhanced IMBH search sensitivity on high mass BBH systems with highly eccentric orbits (simulation set iii).

\begin{figure}[h!]
    \centering
    \includegraphics[width=\linewidth]{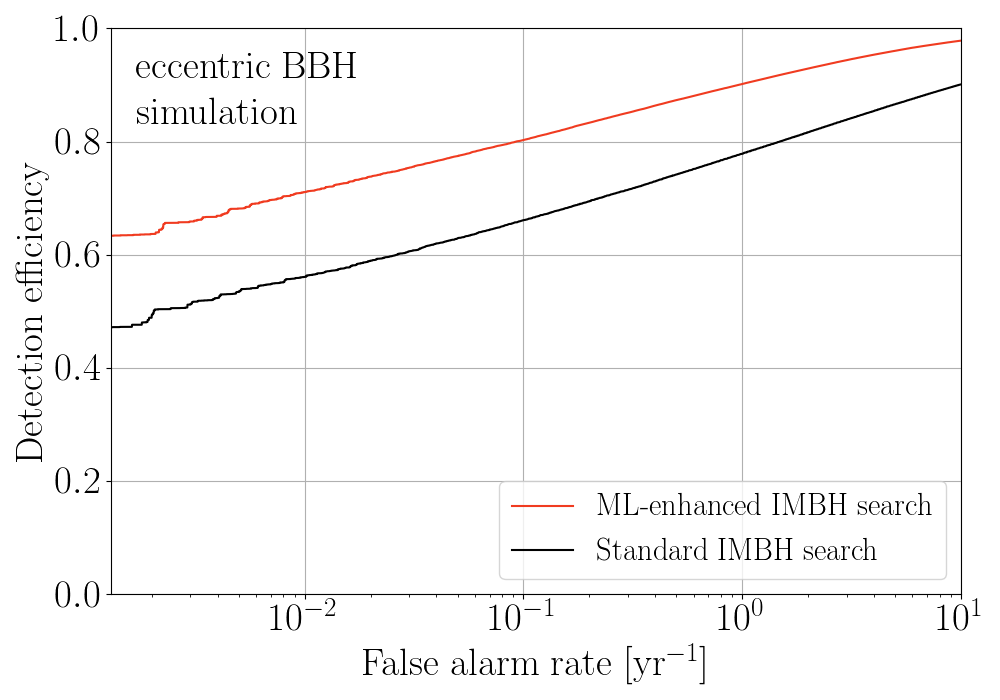}
    \caption{Detection efficiency vs FAR for a high mass, eccentric BBH simulation set recovered with the cWB IMBH configuration. The red curve represents the sensitivity of the ML-enhanced cWB search, and the black curve represents the sensitivity of the standard cWB search.}
    \label{fig:roc_ebbh}
\end{figure}

\begin{figure}[h!]
    \centering
    \includegraphics[width=\linewidth]{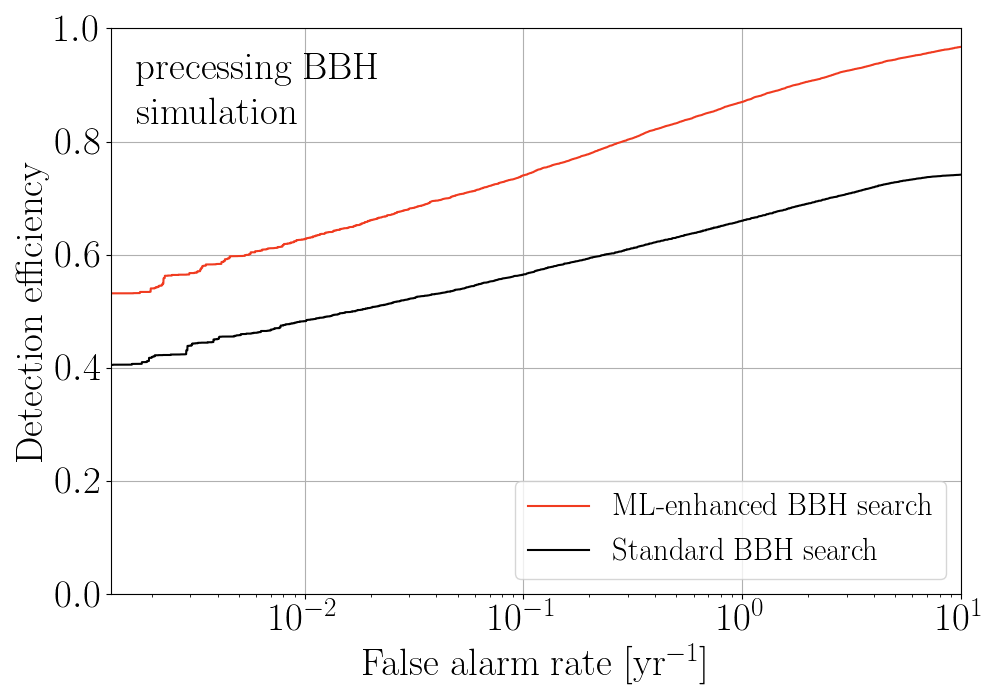}
    \caption{Detection efficiency vs FAR for a precessing BBH simulation set recovered with the cWB BBH configuration. The red curve represents the sensitivity of the ML-enhanced cWB search, and the black curve represents the sensitivity of the standard cWB search.}
    \label{fig:roc_prec}
\end{figure}

In Fig.~\ref{fig:roc_ebbh}, we observe that the ML-enhanced search is more sensitive to the eccentric BBH mergers despite the ML model being trained only on the quasi-circular IMBH waveforms.

A similar test was conducted using precessing BBH systems (simulation set iv).
The ML model is trained on simulated waveforms with the (2,2) harmonic mode, aligned spins, and low mass ratio (1~-~1/4).
The testing set consists of simulated waveforms with higher-order modes, precessing spins, and a higher mass ratio (1~-~1/20).
Fig.~\ref{fig:roc_prec} shows improved sensitivity of the ML-enhanced BBH search as compared to the standard BBH search. 

These results exhibit that the  ML-enhanced search does not compromise the robustness of the standard search but rather improves the sensitivity of the search for BBH signals lying outside the training simulation set.

\subsection{HL vs HLV}\label{subsec:HLvsHLV}
The Advanced Virgo detector joined the GW detector network in the later part of the O2 run. The addition of the Advanced Virgo detector to the existing two detector network (HL) allowed more precise sky localization of the GW events. While adding a third detector is ideally expected to improve the sensitivity, this is not usually the case with the HL and HLV networks. 
%The standard cWB setup requires re-tuning for each detector network and observing run. 
With their common orientation, the LIGO detectors select a well-defined GW polarization state; that feature is exploited by cWB to mitigate their glitches. In contrast, the Virgo detector orientation differs considerably
from that of the LIGO detectors so that glitches in Virgo data cannot be mitigated as efficiently by cWB. Therefore,
the 3-fold network including Virgo, with its current sensitivity level, is usually less sensitive than the  HL network.
\begin{table}[bht]
    \centering
    \setlength{\tabcolsep}{2.2pt}
    
    \begin{tabular}{lcccc}
        \hline
        \hline
        
        Network & Injections & \multicolumn{2}{c}{Recovered} \\
         & & FAR $<\,1\,\mathrm{yr}^{-1}$ & FAR $<\,0.01\,\mathrm{yr}^{-1}$\\
        %& & Total & FAR $<\,1\,\mathrm{yr}^{-1}$ & FAR $<\,10\,\mathrm{yr}^{-1}$ & FAR $<\,100\,\mathrm{yr}^{-1}$\\
        \hline
        \hline
        %HL & $13408$ & $4610$ & $4362$\\
        %HLV & $6943$ & $2400$ & $2092$\\
        
        %HL & $5628$ & $1922$ & $1818$ & $1600$ & $1371$\\
        %HLV & $5628$ & $1939$ & $1688$ & $1451$ & $1221$\\
        %HL+HLV & $5628$ & $2127$ & $1873$ & $1627$ & $1336$ \\
        
        HL & $5628$ &  $1818$ & $1371$\\
        HLV & $5628$ & $1688$ & $1221$\\
        HL+HLV & $5628$ & $1873$ & $1336$ \\
        
        \hline
        \hline
    \end{tabular}
    \caption{Comparison of the number of recovered simulations by the ML-enhanced cWB search at FAR $<\,1\,\mathrm{yr}^{-1}$ and FAR $<\,0.01\,\mathrm{yr}^{-1}$ with respect to the total number of common injections made in limited data (1/33rd of the O3 run) for both the HL and HLV detector networks.
    }
    \label{tab:4}
\end{table}
\begin{figure}[h!]
    \centering
    \includegraphics[width=\linewidth]{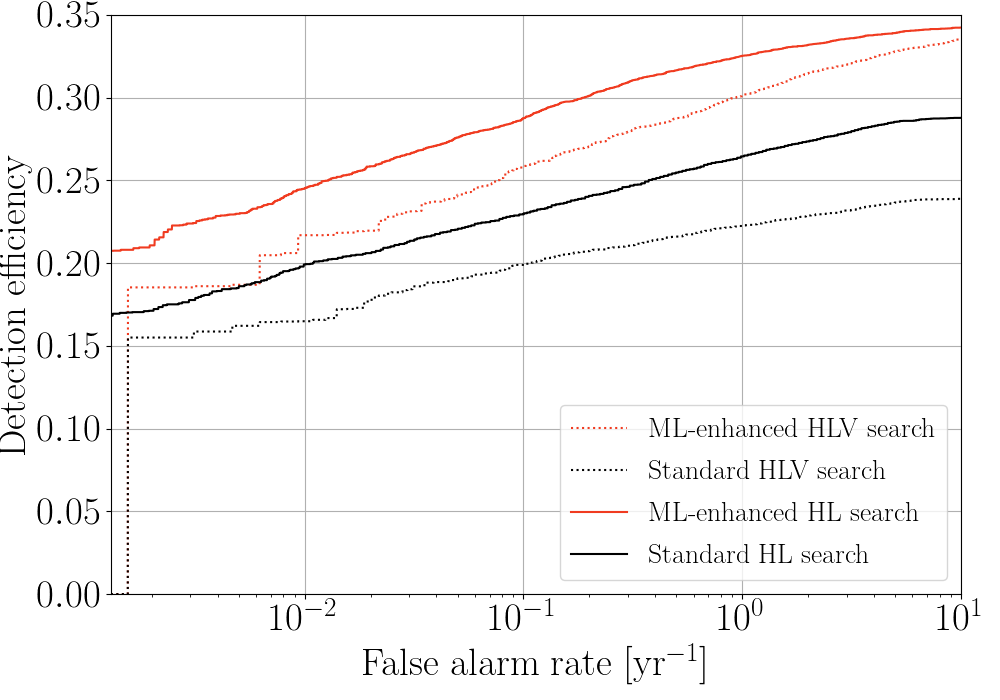}
    \caption{The detection efficiency curves for HL and HLV networks calculated w.r.t. the total number of injections. The solid lines correspond to the HL network, whereas the dotted lines correspond to the HLV network. The blue curves denote the performance of the standard cWB pipeline without ML, and the red curves denote the performance of the ML-enhanced cWB search. The steep drop of detection efficiency curves for the HLV network at low FAR is because of the limited background data ($\sim 640$ years) available for testing.}
    \label{fig:HL_HLV}
\end{figure}
We test the ML-enhanced cWB search on limited data of the HLV network with the BBH search configuration.
%The ML-enhanced cWB search automates the production a continuous ranking criteria for signal and noise classification for any detector network configuration and observing run, while at the same time improves the search sensitivity. 
We train a model by using the accumulated background data and recovered simulations for the HLV network, on the limited data with live time interval - GPS:1241011102.0 to GPS:1242485126.0 which corresponds to 1/33rd of the O3 run (taken from the first half of O3). For the HLV network, we use $550$ years of background data for training. We end up with $\sim 640$ background years of the HLV network 
%, and $\sim 9260$ background years for HL network, 
for testing the performance of the ML model. For the HL network, we use the BBH configuration trained model on the limited data. This exercise considers the total number of common injections made in the limited data live time for both the detector networks and compares their performance based on the recovered simulations at a given FAR as reported in Tab.~\ref{tab:4}.

In Fig.~\ref{fig:HL_HLV}, the detection efficiency is calculated as the ratio of the number of recovered simulations below a given FAR threshold to the total number of injections made in the limited data live time. We observe that the ML-enhanced cWB search improves the sensitivity for both HL and HLV networks. %The HLV network recovers $\sim 1\,\%$ more simulations from the common set of signal injections as compared to the HL network. However, 
The HL network shows better performance and recovers $\sim 7.7\,\%$ more simulated events at FAR $<\,1\,\mathrm{yr}^{-1}$ as compared to the HLV network. In order to assess the performances of these two networks, we combine the HL and HLV network search results by applying a trial factor to the significance of events to check the HL+HLV network performance. We find that the joint search HL+HLV network recovers $\sim 3.0\,\%$ more simulations at FAR $<\,1\,\mathrm{yr}^{-1}$ than the HL network, and $\sim 11.0\,\%$ more simulations at FAR $<\,1\,\mathrm{yr}^{-1}$ than the HLV network. While adding Virgo to the HL network allows us to detect more events, the addition also increases the detector network's background noise, %which leads us to lose
leading to a lower sensitivity at higher significance. For example, the HL network recovers $\sim 2.6\,\%$ more simulations at FAR $<\,0.01\,\mathrm{yr}^{-1}$ than the HL+HLV network. %The reason for the suboptimality of the HLV network is that the Virgo detector is less sensitive than the LIGO detectors. 
With the current cWB implementation and with Virgo lower sensitivity with respect to the LIGO detectors, it is still not beneficial to use the 3-fold network for detection purposes. However, an improvement on the algorithmic side and/or on Virgo sensitivity, as well as the extension to a fourth detector, such as KAGRA, may soon change this condition. 
%The Virgo data brings more noise into the detector network and results in the loss of significance for detected events instead of the two detector network. 
%As a result, it is not beneficial to use the three detector network, or the combination HL+HLV, when the two detector network is available.  
As a result, %it is not beneficial to use the 3-fold detector network, and instead 
we utilize the best available (most sensitive) 2-fold detector network.

\section{Conclusions}
\label{sec:conclusions}
In this paper, we have successfully utilized the ML-enhanced cWB pipeline to search for BBH mergers in the third observing run (O3) of the Advanced LIGO-Virgo. The ML-enhanced cWB search consistently improves the search sensitivities, as previously presented in Ref.~\cite{MLPaper}. For simulated events with FAR $<1\,\mathrm{yr}^{-1}$, the detection efficiency is improved by $18\, \%$ for both the stellar-mass BBH search configuration and the IMBH search configuration. We recover with equivalent or higher significance, all the 15+10 (O3a+O3b) BBH events reported in the third GW Transient Catalog~\cite{GWTC3} that were previously detected by the standard cWB pipeline. 
%The IMBH event GW190521 detected by the ML-enhanced cWB search is two times more significant than the standard cWB search. 
The FAR assigned to the IMBH event GW190521 detected in the ML-enhanced cWB search is limited by the amount of available accumulated background data, and it is two times more significant than the standard cWB search. Additionally, 5 GW events were detected by the ML-enhanced cWB search that were previously missed by the standard cWB search but detected by other search pipelines. Moreover, we detect three cWB-only candidate events with FAR $<1\,\mathrm{yr}^{-1}$ that are not present in the GW Transient Catalogs: 190904\_174910, 190421\_012633, and 191223\_151007. The event 190904\_174910 is found to be consistent with the instrumental noise produced around the 60Hz line by the up-conversion of low-frequency seismic noise. The other two events, 190421\_012633 and 191223\_151007 are marginal IMBH candidates with their source frame total masses $> 100\, M_{\odot}$ lying in the IMBH range. We find broad agreement between the inferred source frame total masses and the cWB estimates $M_{\mathrm{cwb}}$ for these two marginal events. Significant overlap in the 90\% credible sky localization area estimated by cWB and given by the marginal posteriors is observed. While we have not identified any possible instrumental noise origin for these two events, the significance is insufficient to claim confident GW detection. 
%; GW190513, GW190707, GW190728, GW190929, and GW191127 with FAR $<1\,\mathrm{yr}^{-1}$. 

%We report three cWB-only marginal events out of which 

We also test the robustness of the ML implementation by training the ML models on quasi-circular orbit binaries data and testing it on eccentric orbit binaries data, and precessing BBH data. We conclude that the ML-enhanced cWB search does not adversely affect the model independent nature of the cWB pipeline and improves the sensitivity of the pipeline to signals outside the training set.

Lastly, we perform a detector network comparison test, where we establish that the ML-enhanced cWB search enhances the search sensitivity for different detector networks. However, we also note that the HL detector network recovers $\sim 7.7\,\%$ more simulated events at FAR $<1\,\mathrm{yr}^{-1}$ than the HLV network. One of the possible reasons for this is that the Virgo detector brings in more noise to the detector network, preventing improvement in the significance of the detected events.

As a future outlook, the ML-enhanced cWB pipeline will be used in the upcoming fourth observing run of LIGO, and further studies are being done to include this method in the low latency cWB search.

\begin{acknowledgments}

This research has made use of data, software, and/or web tools obtained from the Gravitational Wave Open Science Center, a service of LIGO Laboratory, the LIGO Scientific Collaboration, and the Virgo Collaboration. This work was supported by the NSF Grant No. PHY 1806165 and PHY 2110060. We gratefully acknowledge the support of LIGO and Virgo for the provision of computational resources. I.~B. acknowledges support by the NSF Grant No. PHY 1911796, the Alfred P. Sloan Foundation, and by the University of Florida. 
%We thank Ik Siong Heng, Erik Katsavounidis, Peter Shawhan, and Michele Zanolin for their continued participation and effort in reviewing cWB analyses.
We acknowledge the use of open source Python packages including \textsc{NumPy}~\cite{numpy}, \textsc{Pandas}~\cite{pandas}, \textsc{Matplotlib}~\cite{matplotlib}, and \textsc{scikit-learn}~\cite{scikit-learn}.

\end{acknowledgments}

%%%%%%%%%%%%%%%%%%%%%%%%%%%%%%%%%%%%%%%%%%%%%%%%

\appendix
\section{XGBoost model training updates \label{XGB}}
In this section, we discuss all the updates made to the existing ML method as described in Ref.~\cite{MLPaper}. In order to account for the different detector networks, we replaced the summary statistic that described the quality of the likelihood fit ($E_\mathrm{c}/L$) with ($S_\mathrm{0}/L$) for the two detector network (HL), and with ($S_\mathrm{0}/L$ and $S_\mathrm{1}/L$) for the three detector network (HLV). The summary statistics gives the ratio of the square of the SNR of the reconstructed waveform for each detector ($S_\mathrm{0}$, $S_\mathrm{1}$) to the likelihood ($L$). This allows us to increase the sensitivity of the ML algorithm to different detector networks. The detailed list of selected summary statistics and their definitions can be found in the Appendix of Ref.~\cite{MLPaper}. 

\begin{table}[bht]
    \centering
    \begin{tabular}{cc}
        \hline
        \hline
        XGBoost hyper-parameter & entry \\
        \hline
        \hline
        \texttt{objective} & \texttt{binary:logistic}  \\
        \texttt{tree\_method} & \texttt{hist} \\
        \texttt{grow\_policy} & \texttt{lossguide} \\
        \texttt{n\_estimators} & 20,000$\dagger$ \\
        \texttt{max\_depth} & 13 \\
        \texttt{learning\_rate} & 0.03 \\
        \texttt{min\_child\_weight} & 10.0 \\
        \texttt{colsample\_bytree} & 1.0 \\
        \texttt{subsample} & 0.4, \textbf{0.6}, 0.8 \\
        \texttt{gamma} & \textbf{2.0}, 5.0, 10.0 \\
        \hline
        \hline
    \end{tabular}
    \caption{Entries for XGBoost hyper-parameters. $\dagger$: \texttt{n\_estimators} is optimized using early stopping, where the training stops when the validation score stops improving. Bold entries indicate the optimal choice.}
    \label{tab:4}
\end{table}

The XGBoost hyper-parameters, listed in Table.\ref{tab:4}, were carefully selected to avoid over-fitting and for optimal performance with O3a data for both the BBH and IMBH search configurations. In this study, we first start by setting all XGBoost hyper-parameters to be the same as used for O1 and O2 analysis. We then select \textit{a priori}, the \texttt{max\_depth} to $13$ as we use $13$ cWB summary statistics in the list of input features. We also set the \texttt{colsample\_bytree} to $1$, \textit{a priori}, which ensures that all the features are used while building each tree in the ensemble, during training. Then we perform a short grid search over 9 combinations of XGBoost hyper-parameters with respect to the precision-recall area under the curve (AUC PR) metric~\cite{Davis2006}, and fix the \texttt{subsample} and \texttt{gamma} as reported in Tab.~\ref{tab:4}. The total number of trees generated is optimized by using a method known as \textit{early stopping}, where a small fraction of the training data set is set aside for validation, and the training ends when the validation score with respect to AUC PR metric stops improving with additional trees. This prevents overfitting of the XGBoost algorithm to the training data.

\begin{figure}[h!]
    \centering
    \includegraphics[width=\linewidth]{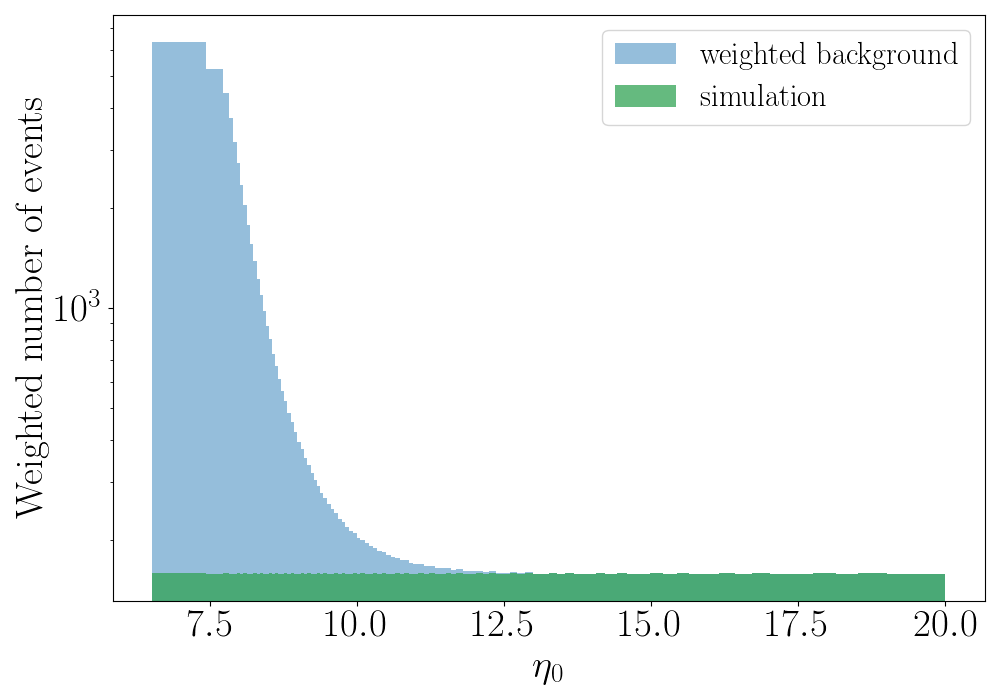}
    \caption{Weighted event distribution as a function of $\eta_\mathrm{0}$ for O3a run BBH search configuration data. The weight options are set to $q=5$ and $A=40$.}
    \label{fig:weight}
\end{figure}
The ML algorithm is computationally efficient and the entire training and testing procedure is completed within minutes, using one CPU core. In order to get rid of excess background with minimal or no loss of simulated events, we employ a cut-off on the $\eta_\mathrm{0}$ statistic with the threshold set at $\eta_\mathrm{0}>6.5$. Before we use this summary statistic as an input feature for XGBoost training, we cap the statistic at $\eta_\mathrm{0} = 20$ (which was earlier set to $8$ because of the previous $\eta_\mathrm{0}$ definition in the paper~\cite{MLPaper}) such that any event with a higher $\eta_\mathrm{0}$ value is assigned a value of $20$. The capping prevents the algorithm from being affected by high SNR background events, which fall steeply with the increase in $\eta_\mathrm{0}$.

In addition to using $\eta_\mathrm{0}$ as an input feature for XGBoost, we also apply a $\eta_\mathrm{0}$ dependent sample weight on the background events to minimize the importance given by the algorithm to low SNR glitches. A standardized sample weight has been proposed in this work as compared to the one reported in Ref.~\cite{MLPaper}. All the simulation events are assigned a sample weight of $1$. In contrast, for the noise events, we first divide simulation events in the interval $6.5<\eta_\mathrm{0}<20$ into $nbins = 100$ percentile bins such that the number of simulation events in each bin is the same. Then, we apply the sample weight per bin as follows:

\begin{equation}\label{eq:3}
    w_{\mathrm{B}}(i) = \frac{N_\mathrm{S}(i)}{N_\mathrm{B}(i)}\, e^{ln(A)\left(1-\frac{i}{nbins}\right)^q} \, ,
\end{equation}
where, $i = {1,2, ..., nbins}$ is a given bin, $N_\mathrm{S}(i)$ and $N_\mathrm{B}(i)$ are the number of simulation and background events in the $i^{\mathrm{th}}$ bin. ($q,\, A$) are weight options where $A$ is called the balance parameter $A=\frac{N_\mathrm{S}(1)}{N_\mathrm{B}(1)}$ is the class balance ($N_\mathrm{S}/N_\mathrm{B}$) for the first bin at $\eta_\mathrm{0} \geq 6.5$, $q$ is called the slope parameter that controls the rate of change of weighted background distribution. For all events with $\eta_\mathrm{0} \geq 20$, the number of simulation events is re-sampled to match the number of background events to have a perfect class balance ($N_\mathrm{S}/N_\mathrm{B} = 1$). For a given combination of ($q,\,A$) values, we can achieve any monotonic distribution of our choice, starting with a class balance of $A$ at $\eta_\mathrm{0}=6.5$ and ending with a perfect class balance at $\eta_\mathrm{0}=20$. In this study, the weight options were fixed to be $q=5$ and $A=40$, resulting in a weighted distribution as shown in Fig.~\ref{fig:weight}. This weight distribution enables the ML model to differentiate high SNR events into noise and signal, while at the same time being sensitive to low SNR events.

%\newpage
\bibliography{paper.bib}

\end{document}